%% file: main.tex
\documentclass{article}

\usepackage{natbib} 
\usepackage{mathtools} 
\usepackage{booktabs} 
\usepackage{tikz} 
\usepackage{graphicx}
\usepackage{algorithm2e}
  \SetKwInOut{Input}{Input}
  \SetKwInOut{Output}{Output}

\usepackage{amsmath}
\usepackage{hyperref}
\hypersetup{colorlinks,allcolors=black}

\newcommand{\footremember}[2]{%
  \footnote{#2}
    \newcounter{#1}
    \setcounter{#1}{\value{footnote}}%
}
\newcommand{\footrecall}[1]{%
    \footnotemark[\value{#1}]%
}

\title{Bayesian Model Averaging of Chain Event Graphs for Robust Explanatory Modelling}

\author{Peter Strong \footremember{complexity}{Centre for Complexity Science, University of Warwick, Coventry CV4 7AL, UK}\footremember{turing}{The Alan Turing Institute, British Library, 96 Euston Road, London NW1 2DB, UK} \\ \href{mailto:P.R.Strong@warwick.ac.uk}{P.R.Strong@warwick.ac.uk} 
\and Jim Q Smith \footremember{stats}{Department of Statistics, University of Warwick, Coventry CV4 7AL, UK}\footrecall{turing} \\ \href{mailto:J.Q.Smith@warwick.ac.uk}{J.Q.Smith@warwick.ac.uk}} 
\begin{document}
\maketitle

\begin{abstract}
Chain Event Graphs (CEGs) are a widely applicable class of probabilistic graphical model that can represent context-specific independence statements and asymmetric unfoldings of events in an easily interpretable way. Existing model selection literature on CEGs has largely focused on obtaining the \textit{maximum a posteriori} (MAP) CEG. However, MAP selection is well-known to ignore model uncertainty. Here, we explore the use of Bayesian model averaging over this class. We demonstrate how this approach can quantify model uncertainty and leads to more robust inference by identifying shared features across multiple high-scoring models. Because the space of possible CEGs is huge, scoring models exhaustively for model averaging in all but small problems is prohibitive. However, we provide a simple modification of an existing model selection algorithm, that samples the model space, to illustrate the efficacy of Bayesian model averaging compared to more standard MAP modelling.

\end{abstract}
\input{1_introduction}
\input{2_Bayesian_learning_of_Chain_Event_Graphs}

\input{3_Model_averaging_of_Chain_Event_Graphs}
\input{4_Methodology}
\input{5_Results}
\input{6_Discussion}

\section*{Acknowledgements}
Peter Strong was supported by the Engineering and Physical Sciences Research Council and the Medical Research Council grant EP/L015374/1. Jim Q. Smith was funded by the EPSRC [grant number EP/K03 9628/1]. We would like to thank Aditi Shenvi for her valuable comments.

\bibliography{ref}

\clearpage

\appendix

\end{document}

%% file: 1_introduction.tex
\section{Introduction}\label{sec:intro}
Chain Event Graphs (CEGs) are a class of interpretable graphical models able to represent asymmetric processes on discrete data, they include discrete context-specific Bayesian Networks (BNs) as a special case. This generalisations means CEGs can represent context-specific independence statements and domains where processes can unfold in diverse ways; the latter are called \textit{non-stratified} CEGs. These properties make CEGs widely applicable and have been used in various domains including educational studies \citep{freeman_smith_2011}, policing \citep{policing}, public health \citep{shenvi2019bayesian} and migration studies \citep{strong2021bayesian}.

CEGs are constructed from event trees by setting up a staging on the non-leaf vertices, the \textit{situations}, of the tree. Two situations are defined to be in the same \textit{stage} when their outgoing edges have the same conditional transition probabilities and share the same real-world meaning. The staging of the tree uniquely defines a \textit{staged tree} and from the set of staged trees there is a bijection to the set of CEGs \citep{smith2008conditional}. A CEG provides a compact representations of complex independence statements, with each CEG for a given tree providing a different explanation of the underlying process. Therefore, for a given event tree, the model selection depends on deciding between independence statements through the potential ways of staging a tree. 

Predominantly, score-based methods have been used to select CEG models, including: Hierarchical Agglomerative  Clustering (HAC) \citep{freeman_smith_2011}, a greedy search algorithm; and dynamic programming \citep{cowell2014causal}, an exhaustive search of the model space which is only possible for small models. K-means has also been used \citep{silander2013dynamic}; however its MAP estimate typically scores lower than the ones found by HAC. 

A single selected CEG, such as the MAP model, often provides helpful insights into the underlying data generating process when that candidate has a high posterior probability. However, when this is not the case, focusing only on this model will lead to overconfident and sometimes spurious inferences. This occurs when there are many high-scoring models with non-negligible probabilities: a phenomenon that is common if the size of the model space dwarfs the number of data points, this is a typical scenario in all but the simplest of settings including for CEGs \citep{k-best}. More recently, non-Bayesian approaches to learning the staging structure based on clustering have been used \citep{JSSv102i06}. However, these methods solely focus on obtaining a single CEG that maximises some score, so are susceptible to the same problems as those outlined above.

Especially when -- as is the case for CEGs-- each scored model has an associated explanation to accompany it, Bayesian model averaging provides an excellent method in which to measure the robustness of explanations to model uncertainty. When explanations are shared across many high-scoring models, then inferences are more secure than when such explanations only apply to the single, highest-scoring model. Of course, except in small model spaces, it is often impossible to average over all models within the class as the number of possible CEGs grows \citep{silander2013dynamic}. To address this issue, Bayesian model averaging is performed over a subset of the model class to approximate the complete, more technically sound, Bayesian model average. Examples of such methods include using k-best models \citep{k-best} and Occam's window \citep{window} giving a subset of the model class. 

Of course, except in small model spaces it is often impossible to even score all models within the class: here the number of possible CEGs grows super-exponentially \cite{silander2013dynamic}. Therefore, instead of scoring every model in the space, we propose the use of a sampler in order to obtain a set of high-scoring models for which to apply Occam's window. This gives an approximation for the true Bayesian model averaging which enables us to better understand how robust our inferences might be to the model selection process.

In the next section, we review the existing literature on model selection of CEGs, the literature on model averaging and how this relates to CEGs. In Section \ref{Methodology}, we describe the simple algorithm we used to search our model space to search a space of CEGs and explore how the explanations embedded in competingly good CEGs can be compared and combined. In Section \ref{Example}, we illustrate our example of our proposed methodology and explain its benefits. The paper ends with a short discussion.

%% file: 2_Bayesian_learning_of_Chain_Event_Graphs.tex
\section{Bayesian learning of CEGs}
\label{CEGs}
\subsection{Notation and assumptions}
Any given probability tree, $\mathcal{T}$, can be decomposed as a set of \textit{florets}. Florets are simple sub-trees of depth 1 rooted at the situations of the tree. A \textit{staged tree} is a probability tree with a staging displayed through colours. The colours partition the situations such that if two situations are coloured the same, they have the same probability distribution over their outgoing edges. Each distinct staging will give rise to a different model in our model space, $M_k \in\mathcal{S}$,  $k=1,\dots,K$, where $K$, the number of staged tree, is given by a product of Bell numbers that grow super-exponentially \citep{silander2013dynamic}. For example, given a stratified staged tree with 5 binary events, there are approximately $1.3 \times 10^{15}$ possible models. 

In practice, for logical reasons, we can often restrict model selection over those CEGs that could make sense. For example, two situations with different numbers of outgoing edges cannot be in the same stage. Less obviously, in any given context, for two edge probabilities to be assumed equal, we must be able to associate the meaning of these edges in some way. Therefore when performing model selection, it is important, both from a modelling and a computational point of view, to restrict the search space so that only models that make sense within a given context are traversed and scored. Clearly, the choice of this restricted space can depend strongly on the domain. 

To restrict the set of models we are considering, we use the concept of a \textit{hyperstage} \citep{collazo2017thesis}. A hyperstage, $\pmb{H}$, is a collection of sets, \textit{hypersets}, containing the situations of an event tree $\mathcal{T}$, which we allow to be in the same stage. We will make the simplification that the hyperstage corresponds to a partition of the set of situations where, before any data analysis has taken place, situations within the same set could be plausibly seen as predictively equivalent, were data to support this \textit{a posteriori}. This simplifies model selection as the staging of each set in the hyperstage is independent. Therefore, we can model fit to each hyperset separately. We note, for example, in all context-specific BNs, we implicitly assume that conditional probabilities can only be hypothesised as being the same when the situations defined by their parents involve the same variables \citep{CEG_book}. 

\subsection{Conjugate learning and model selection}

As in \cite{CEG_book}, we can obtain our posterior probabilities for the probability tree, through the use of a product Dirichlet-Multinomial distribution to perform a conjugate prior to posterior analysis. For simplicity, the priors for our Dirichlet distributions are calculated by choosing an effective sample size, $\bar{\pmb{\alpha}}$, which assumes \textit{a priori} that all florets in the tree are uniformly distributed over. 

The model selection algorithm used for Bayesian model averaging in this paper is based on the HAC algorithm; a greedy search algorithm that begins with each situation as a separate stage, then merges stages that provide the largest improvement in the model likelihood. This is done using \textit{one-nested} CEGs, CEGs which can be obtained from others by merging two stages. This enables fast evaluation of the comparison as the log Bayes factor (BF) of one-nested CEGs can be calculated by only considering situations in which their stagings are different \citep{freeman_smith_2011}.

%% file: 3_Model_averaging_of_Chain_Event_Graphs.tex
\section{CEG Bayesian Model averaging}
\label{model_averaging}
\subsection{Bayesian Model Averaging}
This section begins by providing a brief review of Bayesian model averaging, as in \cite{window}.
When the focus of our inference is on $\Upsilon$ given data $y$ and models $M_k$, $k=1,\dots,K$,
\begin{equation}
    p(\Upsilon|y)=\sum_{k=1}^K p(\Upsilon|M_k)p(M_k|y).
\end{equation}
This shows that the prediction a weighted average of each of the $K$ competing models. Here, $p(\Upsilon|M_k)$ is the posterior probability of model $M_k$ and $p(M_k|y)$,
\begin{equation}
    p(M_k|y)=\frac{p(y|M_k)p(M_k)}{\sum_{i=1}^Kp(y|M_i)p(M_i)}.
    \label{model_prob}
\end{equation}

For each model, $p(M_k|y)$ captures its posterior uncertainty.

We can represent the posterior odds of two models as follows
\begin{equation}
    \frac{p(M_k|y)}{p(M_l|y)}=\frac{p(y|M_k)}{p(y|M_l)}\times \frac{p(M_k)}{p(M_l)}.
\end{equation}
The BF is $BF_{k,l}=\frac{p(y|M_k)}{p(y|M_l)}$. Therefore we can represent Equation \ref{model_prob} as follows,
\begin{equation}
    p(M_k|y)=\frac{BF_{k,1}p(M_k)}{\sum_{i=1}^K BF_{i,1}p(M_i)}.
\end{equation}
 As we set a uniform prior $p(M_i)=p(M_j)$ for all $i, j \in \{1,\dots,K\}$, therefore
 \begin{equation}
    p(M_k|y)=\frac{BF_{k,1}}{\sum_{i=1}^K BF_{i,1}}.
    \label{BF ratio}
\end{equation}
 
This means that we are able to work out the model weighting based off all of the BFs against a first model.

\subsection{Occam's Window}
Performing model averaging over all models is often an intractable problem due to the number of potential models. Therefore, when performing model averaging, it is necessary to reduce the number of models considered; Occam's window helps us to keep in the frame only the most supported. This is based on two steps: first, we remove any of our choice models that are $\beta$ times less likely than the best performing model \footnote{A standard choice of $\beta$ is 20 \citep{MCMC3}.}.

\begin{equation}
    \mathcal{S}'=\bigg \{M_k :M_k \in \mathcal{S} \land \frac{\max_{M_l \in \mathcal{S}} p(M_l|y)}{p(M_k|y)} <\beta \bigg \}
\end{equation}

Secondly, we discard any models for which there exists a nested, more likely model, as with Occam's razor. In Equation \ref{razor} we denote $M_l \subset M_k$ if $M_l$ is a nested model of $M_K$.

\begin{equation}
    \mathcal{R}=\bigg \{M_k : \exists M_l \in \mathcal{S}', M_l \subset M_k \land \frac{p(M_k|y)} {p(M_l|y)}<1 \bigg \}
    \label{razor}
\end{equation}

\begin{equation}
    \mathcal{\hat{S}}=\mathcal{S'}\backslash \mathcal{R}
\end{equation}

This leaves a set of models, $\mathcal{\hat{S}}$, which we will refer to as the \textit{well-performing} models. When dealing with interpretable models, Occam's window is not just an approximation. It also effectively enables us to focus on good explanatory models and discard the rest. This is vital when there might be many poorly fitting models in the space \textit{a priori}; although none of these explains the process well, the residual probability on these remains large after sampling, which blurs the posterior image until we are able to gather enormous amounts of data. This will be the case when, for example, we do not have the time or expertise to forensically set priors on models that \textit{a priori} should be assigned a small probability.

\subsection{Model averaging CEGs}
When setting a uniform prior over a set of models, it is important that no models are represented twice.
While the operations that define the possible ways to transverse the equivalence class of CEGs is known \citep{gorgen18}, the cardinally of each equivalence class remains an open research problem. For a fixed tree, there only exists a single model in each equivalence class. Therefore, in this paper, the probability tree $\mathcal{T}$, on which the CEG is based is fixed \textit{a priori}. 

As the number of possible staged trees grows super-exponentially, calculating the likelihoods for each model in the space of CEGs is an intractable problem. In this paper, we propose the following approach to address this issue. First, sample the model space. Then, apply Occam's window to obtain an approximation of a set of well-performing models. Our approach is to obtain a set of models that are a good approximation of Occam's window. This provides a good narrative, across a few alternatives, that, under certain conditions, can be shown to provide an approximation to the true set of models with highest associated posterior probabilities. Most importantly, this approximation will allow for quantification of model uncertainty, and therefore certainty in the model's independence statements.

%% file: 4_Methodology.tex
\section{Methodology}
\label{Methodology}
In this paper, for clarity, we content ourselves with comparing the performance of model averaging against MAP estimation for searching across explanations using a very simple -- although, as far as we know, novel -- search algorithm. This is because the benefits of model averaging do not depend strongly on this choice, provided the method is able to search in the neighbourhood of high-scoring models. We show that, even with this naive method, our search performs surprisingly well for the purposes of Bayesian model averaging.

This method is a weighted version of the HAC algorithm, weighted hierarchical agglomerative clustering (w-HAC). Instead of being a greedy search algorithm, as in HAC, w-HAC is a randomised algorithm, where the probability of merging is weighted by the relative BFs of the potential mergings. Under w-HAC, the probability of two stages merging is given in Equation \ref{choice}. Here, $BF_{i\oplus j,1}$ is the BF comparing when stages $i$ and $j$ are merged to when they are not. 

\begin{equation}
    p(u_i,u_j) = \frac{BF_{i\oplus j,1}}{\sum_{k,l \in H_i} BF_{k\oplus l,1}}
    \label{choice}
\end{equation}

Due to the stopping criteria for w-HAC, as in HAC -- either stopping if there are no possible situations left to be merged or if none of the potential mergers would increase the $BF$ -- we know that our set of solutions will satisfy a weaker version of Occam's razor: it will not include a less probable model that is more complicated than a one-nested simpler model. This is because, due to the latter condition, if a simpler nested model existed, there would be a merged model with a positive BF. Therefore, a merging would occur.

For explanatory modelling, it is important to obtain the most likely model. For CEGs, a method, other than exhaustive search, for finding the true MAP is still an open research question, with HAC currently providing the best estimates. As HAC is simply a greedy version of w-HAC, we hypothesise that the set of models outputted by iterations of w-HAC will contain HAC's MAP estimate alongside other high scoring models as well as potentially avoiding local maxima. We note that, while K-means, for different K values, could be used to give a set of models to perform model averaging on, this would be a poorer choice since its output has been shown to perform worse than HAC \citep{silander2013dynamic}.

\subsection{Independent hyperstage set staging}
As when applying the HAC algorithm, w-HAC can be run on each hyperset in the hyperstage independently. This has two main benefits: first, the different potential stagings of a hyperset can be studied independently; secondly, it allows the prioritisation of computational resources to hypersets with many more elements and therefore many more possibilities of staging. The set of CEGs that we are model averaging over can then be obtained by taking all possible combinations of well-performing stagings of each hyperset.

For hyperset $H_i\in \pmb{H}$, we propose running w-HAC; ${K\times \#(H_i)}$ times. Here $\#H_i$ is the number of elements in hyperset $H_i$ and $K$ is an choice based upon available computational resources. This sets the number of iterations of w-HAC as proportional to the number of elements in the hyperset. We propose this despite the fact that the number of possible stagings, and therefore the search space, rises much faster than the number of situations because if we chose the number of iterations as proportional to the number of stagings, all runs would be focused on the later stages due to the super-exponential growth of the Bell numbers.

%% file: 5_Results.tex
\section{Example}
\label{Example}
\subsection{The Dataset}
\begin{figure}[h!]
\centering
\includegraphics[width=0.73\linewidth]{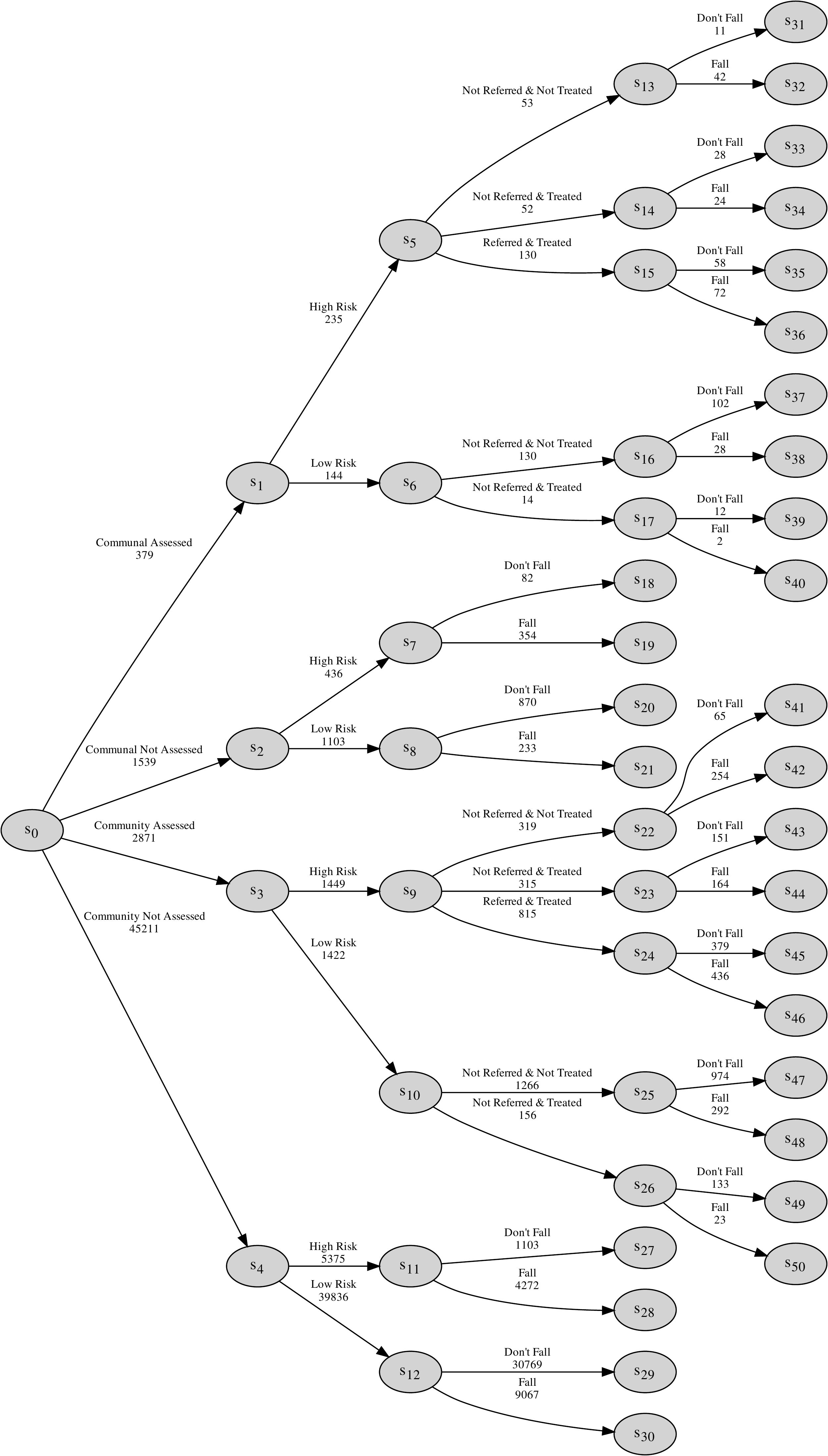}
\caption{Event tree for the simulated falls dataset with the counts for each path.}
\label{nonstrEvent_tree}
\end{figure}

Here, we provide an example of Bayesian model averaging to demonstrate our proposed methodology and show its benefits \footnote{Code for this example can be found here: \url{https://github.com/peterrhysstrong/cegpy_BMA/tree/dev}}. To do this, we work through its application to a simulated falls dataset of 10,000 individuals aged over 65 \citep{shenvi2018modelling}. This dataset was chosen as it is simulated data in which the data-generating CEG is known. The event tree is constructed from the following four variables in the order presented below:
\begin{enumerate}
    \item $X_A$: Individual living situation and whether they have been assessed (Communal Assessed, Communal Not Assessed, Community Assessed, Community Not Assessed)
    \item $X_R$: Level of risk from a fall (High Risk, Low Risk)
    \item $X_{T1}$: If an individual has been referred and treated (Not Referred \& Not Treated, Not Referred \& Treated, Referred \& Treated)
    \item $X_{T2}$: If an individual has been treated (Not Referred \& Not Treated, Not Referred \& Treated)
    \item $X_F$: If a fall happened or not (Fall, Don't Fall)
\end{enumerate}
The event tree describing this unfolding of the events can be seen in Figure \ref{nonstrEvent_tree}. This is a non-stratified event tree because the process can unfold in a variety of ways. For example, for individuals that are not assessed for their risk of falls, it does not make sense to consider the outcome of their assessment. This is a subset of the full dataset containing a fifth of the number of entries chosen to introduce more uncertainty about the presence of different independence statements\footnote{In the full data set there are only two well-performing models.}.

\subsection{The Input}
Here, we will be running w-HAC over each hyperset in the hyperstage. The 5 hypersets in the hyperstage, given in Equation \ref{hyperstage_non_strat}, correspond to a hyperset for each variable. For the other parameters we set $K= 100$, $\beta=20$ and $\bar{\pmb{\alpha}}=4$.

\begin{equation}
\begin{split}
\pmb{H}=&
    \{\{s0\}, \{s1, s2, s3, s4\}, \{s5, s9\}, \{s6, s10\},\\
    & \{s7,  s8,  s11,  s12,  s13,s14,  s15, s16,  s17,  s22,  s23,  s24,s25, s26\}\}
\end{split}
\label{hyperstage_non_strat}
\end{equation}

\subsection{Bayesian Model Averaging}

\begin{figure}[h]
\centering
\includegraphics[width=0.75\linewidth]{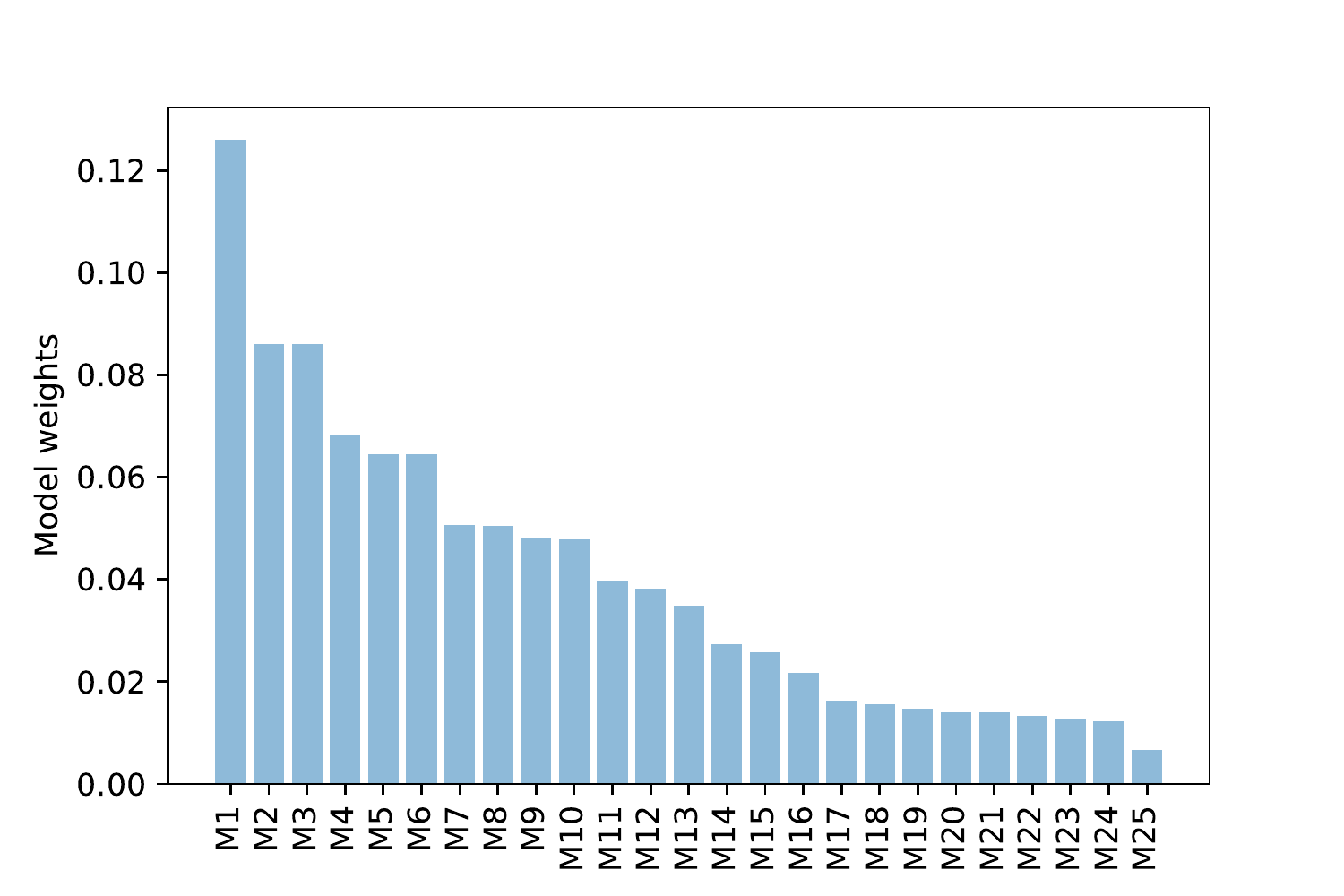} 
\caption{The 25 well-performing models with their model weights given by the ratio of Normalised BFs}
\label{BF_nonstrat}
\end{figure}

For four of the five hypersets within the hyperstage, there was a single well-performing staging. These four hypersets had the following unique well-performing staging:
$\{s0\}, \{s1\}, \{s2\},\\ \{s3\}, \{s4\}, \{s5, s9\}, \{s6, s10\}$. The remaining hyperset within the hyperstage-- $\{s7,  s8,  s11,\\  s12,  s13, s14,  s15, s16,  s17,  s22,  s23,  s24, s25, s26\}$--
had 73 unique stagings outputted from w-HAC; 25 were well-performing. As only one set in the hyperstage has more than one well-performing staging, the model average is over 25 CEGs, which only differ in that one set in the hyperstage. The model weights for each well-performing model are shown in Figure \ref{BF_nonstrat}. 

For the hypersets which have more than one well-performing staging, uncertainty about the staging can be explored by considering the well-performing intersection and union: the intersection and union of all of the stagings of well-performing models. Situations that were in the same stage in the intersection will be in the same stage in all well-performing models; likewise, situations in a different stage in the union will be in a different stage in all well-performing models.
The well-performing union is equal to the hyperset and the well-performing intersection is:\\
$\{\{s8,s12,s25,s26\},\{s7, s11, s22\},\{s23, s24\}, \{s13\}, \{s14\},  \{s15\}, \{s16\}, \{s17\}$\}. 

\begin{figure}[h]
\centering
\includegraphics[width=\linewidth]{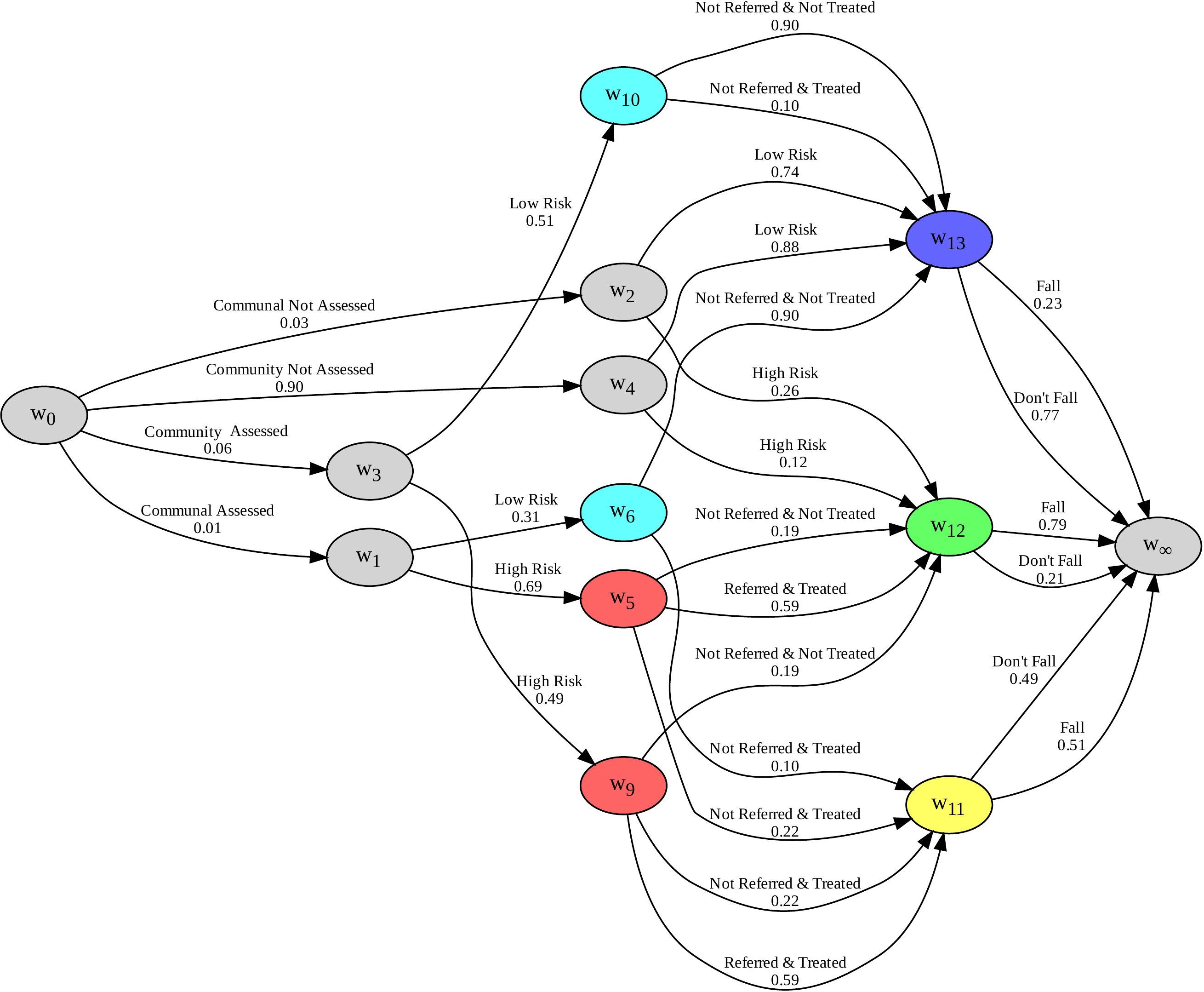}   
\caption{The MAP CEG obtained via HAC for the falls dataset with the mean transition probabilities given along each edge.}
\label{MAP_CEG_nonstrat}
\end{figure}

For comparison, we also ran HAC to obtain its MAP estimate. The model obtained was the same as the highest-weighted model from the output of Bayesian model averaging and can be seen in Figure \ref{MAP_CEG_nonstrat}.

\subsection{Explanation of Results}
In 4 of the 5 hypersets, the well-performing unions and intersections are the same as there is only a single well-performing staging from w-HAC. These four stagings align with the staging in the data-generating process. Three of these hypersets have non-trivial stagings. For these hypersets, the unique, well-performing staging represents the following independence statements, which exist in all of the well-performing models:
\begin{itemize}
    \item The level of risk from a fall is not independent of living situations and assessment status.
    \item The outcome of the assessment for high risk individuals that were assessed is independent of living situation.
    \item The outcome of the assessment for low risk individuals that were assessed is independent of living situation. 
\end{itemize}

For the only hyperset with multiple well-performing stagings, 25 well-performing models exist. Therefore, there are many possible explanations that are a good description of the data. The non-singletons in the well-performing intersection correspond to independence relationships that exist in all of the well-performing models. The singletons in the well-performing intersection are the situations associated with Communal living who were assessed. However, these situations were not singletons in any of the well-performing models, reflected by the well-performing union containing every element in the hyperset. This suggests that, rather than these situations having a distribution over their edges different to any other situation in the hyperset, there is significant uncertainty about how these situations should be staged. As there are only 379 counts for Communally living assessed individuals (Figure \ref{nonstrEvent_tree}) and a weak prior was chosen, it is to be expected that we have little certainty about the staging of these situations.

The HAC MAP estimate of this hyperset is not the data-generating process and the well-performing intersection contains stagings, which are not subsets of the staging of the data generating process. However, comparing against the alternative of considering one model, this provides clear benefit in modelling the uncertainty. 

%% file: 6_Discussion.tex
\section{Discussion}
\label{Discussion}
In this paper, we have provided a methodology that enables a wider class of models, capable of dealing with asymmetric processes, to obtain the benefits of Bayesian model averaging. This class of models can model situations where using BNs would be wholly unsuitable, which significantly increases the applicability of Bayesian model averaging. 
The benefits of applying Bayesian model averaging for CEGs, through using a simple sampling algorithm, are clear: through exploring multiple well-performing models, the robustness of complex independence statements can be quantified by considering them within the set of well-performing models; when multiple well-performing models exist, the explanations shared by them can be extracted. These are significant benefits compared to modelling that involves a single MAP estimate, where there is no quantification of the uncertainty of each independence statement. Although this could be done through diagnostic testing, the Bayesian model averaging approach provides an intuitive solution when diagnostic tests identify many models with high posterior probability. 

Of course, although our naive sampler was sufficient to demonstrate the efficacy of model averaging methods when extracting robust inferences within classes of explainable models, it would be worthwhile to explore more sophisticated methods for extracting high-scoring models. Many alternative sampling algorithms are likely to outperform w-HAC and should ideally satisfy various attractive consistency properties-- such as being able to search the space for highly-separated models -- which could provide very different possible high-scoring explanations of how the data is generated. An additional attractive property would be the ability to scale up to problems with a much larger number of situations. The methodology described in this paper is not specific to our choice of algorithm and is more broadly applicable to any sampling algorithms. This further motivates the need for efficient model selection algorithms for CEGs. Newly available code -- the CEGpy python package \footnote{https://pypi.org/project/cegpy/} and the stagedtree package in R (for stratified CEGs) \citep{JSSv102i06} -- means that the practical efficacy of these can be more easily explored.

%% file: main.bbl
\begin{thebibliography}{15}
\providecommand{\natexlab}[1]{#1}
\providecommand{\url}[1]{\texttt{#1}}
\expandafter\ifx\csname urlstyle\endcsname\relax
  \providecommand{\doi}[1]{doi: #1}\else
  \providecommand{\doi}{doi: \begingroup \urlstyle{rm}\Url}\fi

\bibitem[Bunnin and Smith(2021)]{policing}
F.~Oliver Bunnin and Jim~Q. Smith.
\newblock {A {B}ayesian {H}ierarchical {M}odel for {C}riminal
  {I}nvestigations}.
\newblock \emph{Bayesian Analysis}, 16\penalty0 (1):\penalty0 1--30, 2021.

\bibitem[Carli et~al.(2022)Carli, Leonelli, Riccomagno, and
  Varando]{JSSv102i06}
Federico Carli, Manuele Leonelli, Eva Riccomagno, and Gherardo Varando.
\newblock The r package stagedtrees for structural learning of stratified
  staged trees.
\newblock \emph{Journal of Statistical Software}, 102\penalty0 (6):\penalty0
  1–30, 2022.
\newblock \doi{10.18637/jss.v102.i06}.

\bibitem[Collazo et~al.(2018)Collazo, G{\"o}rgen, and Smith]{CEG_book}
Rodrigo~A. Collazo, Christiane G{\"o}rgen, and Jim~Q Smith.
\newblock \emph{Chain event graphs}.
\newblock {CRC} {P}ress, 2018.

\bibitem[Collazo(2017)]{collazo2017thesis}
Rodrigo~Abrunhosa Collazo.
\newblock \emph{The Dynamic Chain Event Graph}.
\newblock PhD thesis, The {U}niversity of {W}arwick, 2017.

\bibitem[Cowell and Smith(2014)]{cowell2014causal}
Robert~G. Cowell and Jim~Q. Smith.
\newblock Causal discovery through {MAP} selection of stratified chain event
  graphs.
\newblock \emph{Electronic {J}ournal of {S}tatistics}, 8\penalty0 (1):\penalty0
  965--997, 2014.

\bibitem[Fragoso et~al.(2018)Fragoso, Bertoli, and Louzada]{MCMC3}
Tiago~M. Fragoso, Wesley Bertoli, and Francisco Louzada.
\newblock Bayesian {M}odel {A}veraging: {A} {S}ystematic {R}eview and
  {C}onceptual {C}lassification.
\newblock \emph{International Statistical Review}, 86\penalty0 (1):\penalty0
  1--28, 2018.

\bibitem[Freeman and Smith(2011)]{freeman_smith_2011}
Guy Freeman and Jim~Q. Smith.
\newblock {Bayesian MAP model selection of Chain Event Graphs}.
\newblock \emph{Journal of Multivariate Analysis}, 102\penalty0 (7):\penalty0
  1152–1165, 2011.

\bibitem[Görgen and Smith(2018)]{gorgen18}
Christiane Görgen and Jim~Q. Smith.
\newblock {Equivalence classes of staged trees}.
\newblock \emph{Bernoulli}, 24\penalty0 (4A):\penalty0 2676 -- 2692, 2018.
\newblock \doi{10.3150/17-BEJ940}.

\bibitem[Madigan and Raftery(1994)]{window}
David Madigan and Adrian~E. Raftery.
\newblock Model {S}election and {A}ccounting for {M}odel {U}ncertainty in
  {G}raphical {M}odels {U}sing {O}ccam's {W}indow.
\newblock \emph{Journal of the American Statistical Association}, 89\penalty0
  (428):\penalty0 1535--1546, 1994.

\bibitem[Shenvi and Smith(2019)]{shenvi2019bayesian}
Aditi Shenvi and Jim~Q Smith.
\newblock A {B}ayesian dynamic graphical model for recurrent events in public
  health.
\newblock ar{X}iv preprint ar{X}iv:1811.08872, 2019.

\bibitem[Shenvi et~al.(2018)Shenvi, Smith, Walton, and
  Eldridge]{shenvi2018modelling}
Aditi Shenvi, Jim~Q Smith, Robert Walton, and Sandra Eldridge.
\newblock Modelling with non-stratified chain event graphs.
\newblock In \emph{International {C}onference on {B}ayesian {S}tatistics in
  {A}ction}, pages 155--163. Springer, 2018.

\bibitem[Silander and Leong(2013)]{silander2013dynamic}
Tomi Silander and Tze-Yun Leong.
\newblock A dynamic programming algorithm for learning {C}hain {E}vent
  {G}raphs.
\newblock In \emph{International {C}onference on {D}iscovery {S}cience}, pages
  201--216. Springer, 2013.

\bibitem[Smith and Anderson(2008)]{smith2008conditional}
Jim~Q Smith and Paul~E Anderson.
\newblock Conditional independence and chain event graphs.
\newblock \emph{Artificial Intelligence}, 172\penalty0 (1):\penalty0 42--68,
  2008.

\bibitem[Strong et~al.(2021)Strong, McAlpine, and Smith]{strong2021bayesian}
Peter Strong, Alys McAlpine, and Jim~Q. Smith.
\newblock A {B}ayesian {A}nalysis of {M}igration {P}athways using {C}hain
  {E}vent {G}raphs of {A}gent {B}ased {M}odels.
\newblock ar{X}iv preprint. ar{X}iv:2111.04368, 2021.

\bibitem[Tian et~al.(2010)Tian, He, and Ram]{k-best}
Jin Tian, Ru~He, and Lavanya Ram.
\newblock Bayesian {M}odel {A}veraging {U}sing the k-best {B}ayesian {N}etwork
  {S}tructures.
\newblock In \emph{Proceedings of the Twenty-Sixth Conference on Uncertainty in
  Artificial Intelligence}, UAI'10, page 589–597. AUAI Press, 2010.

\end{thebibliography}
